\documentclass[11pt]{article}

\usepackage{amsmath}
\usepackage{amsfonts}
\usepackage{wasysym}
\usepackage{graphicx}

\numberwithin{equation}{section}

\allowdisplaybreaks[3] 

\def\tn{\textnormal}

\begin{document}


\thispagestyle{empty}

\hfill LBNL-61473

\hfill UCB-PTH-06-18

\hfill hep-ph/0608279

\hfill August 24, 2006

\addvspace{45pt}

\begin{center}

\Large{\textbf{``Semi-Realistic" $F$-term Inflation Model Building in Supergravity}}
\\[35pt]
\large{Ben Kain}
\\[10pt]
\textit{Berkeley Center for Theoretical Physics and Department of Physics}
\\ \textit{University of California, Berkeley}
\\ \textit{and}
\\ \textit{Theoretical Physics Group, Lawrence Berkeley National Laboratory}
\\ \textit{Berkeley, CA 94720, USA}
\\[10pt] 
bkain@berkeley.edu
\end{center}

\addvspace{35pt}

\begin{abstract}
\noindent We describe methods for building ``semi-realistic" models of $F$-term inflation.  By semi-realistic we mean that they are built in, and obey the requirements of, ``semi-realistic" particle physics models.  The particle physics models are taken to be effective supergravity theories derived from orbifold compactifications of string theory, and their requirements are taken to be modular invariance, absence of mass terms and stabilization of moduli.  We review the particle physics models, their requirements and tools and methods for building inflation models.
\end{abstract} 

\newpage


\section{Introduction}
Inflation provides answers for many questions concerning the early universe.  This is remarkable given that we do not have a definite model of inflation.  In fact, we do not even know what particle physics model one should attempt inflation model building in.  In some cases this has led to ad hoc proposals for inflaton potentials or inflation model building only loosely based on an underlying particle physics model.  Realistic models of inflation must certainly agree with observation, but they should also emerge from a realistic particle physics model.

As an attempt in this direction we describe methods for building ``semi-realistic" models of $F$-term inflation.  By semi-realistic we mean that they are built in, and obey the requirements of, ``semi-realistic" particle physics models, taken here to be effective supergravity theories derived from orbifold compactifications of string theory.  We consider such effective supergravity theories to be semi-realistic because they have the potential to explain much of our universe in a self consistent way.

This paper is in large part a review of those ideas relevant for inflation model building in the class of supergravity theories we are considering.  In the first part of this paper, making up sections \ref{scalarpots}--\ref{vevs}, we review the effective supergravity theories.  This includes the construction of scalar potentials---with a comprehensive matter content---in two different formalisms, canonical normalization of possible inflatons, string theory requirements the effective supergravity theories should obey and methods for generating VEVs.  This review is intended for the nonspecialist.  In section \ref{modelbuilding} we consider inflation model building.  This includes a method for building inflation models, a review of previous attempts and, by combining the work of these previous attempts, the construction of a small field inflaton potential.  We conclude in section \ref{con}.  

In the remainder of this introduction we briefly review the standard methods for analyzing inflation models \cite{ll, lr}.  In this paper we set the reduced Planck mass, $m_P=1/\sqrt{8\pi G}=2.4\times 10^{18}$ GeV, equal to one: $m_P=1$.  Then, given a scalar potential, $V$, the slow roll parameters are
\begin{equation} \label{srpar}
	\epsilon=\frac{1}{2}\left(\frac{V'}{V}\right)^2, \qquad \eta=\frac{V''}{V}, \qquad \xi^2=\frac{V'V'''}{V^2},
\end{equation}
where a prime denotes differentiation with respect to the inflaton.  Inflation occurs while $\epsilon$, $|\eta| \ll 1$ and is taken to end when one of $\epsilon$, $|\eta|$ is no longer less than one.  The spectral index, $n$, its running, $\alpha=dn/d\ln k$, and the tensor fraction, $r$, are given by
\begin{equation}\label{infeqs}
	n=1-6\epsilon+2\eta, \qquad \alpha=16\epsilon\eta-24\epsilon^2-2\xi^2, \qquad r=16\epsilon.
\end{equation}
Assuming negligible running and tensor fraction, the spectral index has been measured to be \cite{wmap}
\begin{equation}\label{wmap}
	n(\phi_*)=0.95 \pm 0.02,
\end{equation}
where $\phi_*$ is defined to be the value of the inflaton corresponding to this measurement.  The number of efolds from $\phi=\phi_*$ to the end of inflation at $\phi=\phi_e$ is given by
\begin{equation} \label{efoldeq}
	N(\phi_*)=-\int_{\phi_*}^{\phi_e} \frac{V}{V'}\;d\phi,
\end{equation}
with reasonable values being $N(\phi_\star)\approx 50\tn{ -- }60$ \cite{lleach}.  Finally, the COBE normalization requires
\begin{equation}\label{cobe}
	V^{1/4}=\epsilon^{1/4} \cdot 6.6\times 10^{16} \tn{ GeV},
\end{equation}
which is to be evaluated at a very precise scale.  We may take this scale to approximately correspond to $\phi_*$.


\section{Scalar Potentials} \label{scalarpots}
The ``semi-realistic" particle physics models, within which we will consider inflation model building, are effective supergravity theories derived from orbifold compactifications of the weakly coupled heterotic string \cite{dhvw, bl}.  For field content they contain the dilaton, three (diagonal) K\"ahler moduli, untwisted matter fields, twisted matter fields and gaugino condensates (allowing for the possibility of spontaneously breaking supersymmetry) which have been integrated out, inducing a nonperturbative contribution to the superpotential.  In this section we review the construction of the scalar potentials in two different formalisms and discuss some of their differences.

A brief remark on notation: We will often use the word ``superfield," but in an abuse of notation we will always write the lowest (scalar) component.  For example, though we will mention the (anti)chiral superfields $S$, $\bar{S}$, we will always write the scalar fields $s$, $\bar{s}$.

Supergravity theories derived from string theory can be constructed in two dual formalisms: the more common, \textit{chiral superfield formalism} \cite{bl}, wherein a chiral superfield contains the dilaton, $s+\bar{s}$, as the real part of its lowest component, or the \textit{linear superfield formalism} \cite{siegel, bgg}, wherein a linear superfield contains the dilaton, $\ell$, as its lowest component.  We will present the scalar potential in each formalism, but before doing so we consider aspects common to both.

The complete K\"ahler potential is unknown.  We assume, for both the chiral and linear superfield formalisms, that it includes the terms
\begin{equation}\label{kpotgen}
	K\supset-\sum\nolimits_I \ln x_I + \sum\nolimits_A X_A,
\end{equation}
with
\begin{equation}
	x_I=t_I+\bar{t}_I-\sum\nolimits_A|\phi_{AI}|^2, \qquad X_A=\left(\prod\nolimits_I x_I^{-q_I^A}\right) |\phi_A|^2,
\end{equation}
where $t_I$, $I=1,2,3$, are the three (diagonal) K\"ahler moduli fields, $\phi_{AI}$ are the untwisted matter fields with modular weights $q_J^{AI}=\delta_J^I$ and $\phi_A$ are the twisted matter fields with modular weights $q_I^A \geq0$.\footnote{In the literature one also finds modular weights denoted by $n^I_\alpha$, where $n^I_\alpha=-q_I^\alpha$ for $\alpha=AJ$, $A$.\label{fnmw}}  The K\"ahler potential for twisted matter is known only to leading (quadratic) order.  Consequently, twisted matter fields must be assumed small, so that higher order terms are negligible.  No such assumption is required for untwisted matter fields since its K\"ahler potential is known to all orders. In section \ref{modin} we will explain the role of the modular weights and why the requirement of modular invariance leads to the introduction of a Green-Schwarz counterterm \cite{gs},
\begin{equation}\label{vgs}
	V^\tn{GS}=-\sum\nolimits_I b_I\ln x_I+\sum\nolimits_A p_A X_A,
\end{equation}
where the values of $p_A$ are unknown (the values of $b_I$ are given below).  For concreteness we make the plausible assumption $p_A\approx b_I$ \cite{GLM}.  How the Green-Schwarz counterterm is implemented is specific to the formalism, as will be shown below.


\subsection{The Chiral Superfield Formalism} \label{csf}
In the chiral superfield formalism, the K\"ahler potential, for our matter content, is commonly taken to be
\begin{equation}\label{chikpot}
	K=-\ln Y-\sum\nolimits_I \ln x_I + \sum\nolimits_A X_A,
\end{equation}
where
\begin{equation}\label{chiy}
	Y=s+\bar{s}-V^\tn{GS},
\end{equation}
with $V^\tn{GS}$ defined in (\ref{vgs}).  It is conventional in this formalism to write the Green-Schwarz coefficient, $b_I$, as\footnote{Though it is conventional in the chiral superfield formalism for the Green-Schwarz coefficient $\delta_I$ (or $b_I$) to be written with the subscript $I$, most compactifications lead to $I$ independent $\delta_I$.\label{fngs}}
\begin{equation}
	b_I=\frac{\delta^\tn{GS}_I}{8\pi^2}.
\end{equation}
Standard compactifications lead to $\delta^\tn{GS}_I\leq30$ \cite{gs}.  We will refrain from using $\delta_I$ to keep the clutter down in equations.  The field dependence of the superpotential is
\begin{equation}\label{chisp}
	W=W(s,t_I,\phi_{AI},\phi_A).
\end{equation}
Its form will be given in section \ref{sp}.

The scalar potential is made up of the $F$-term and the $D$-term: $V=V_F+V_D$.  $V_F$ is given by
\begin{equation}\label{genpot}
	V_F=e^K\Bigl[K^{m\bar{n}}(W_m+K_m W)(\overline{W}_{\bar{n}}+K_{\bar{n}} \overline{W})-3|W|^2\Bigr],
\end{equation}
where a subscript $m$ refers to a derivative with respect to a chiral superfield, such as $t_I$, and a subscript $\bar{n}$ refers to a derivative with respect to an antichiral superfield, such as $\bar{t}_I$.  $K^{m\bar{n}}$ is the inverse of the K\"ahler metric, $K_{m\bar{n}}$.  Both the K\"ahler metric and its inverse are given in appendix \ref{kmc}.  We will consider $V_D$ in section \ref{dvevs}.  In the absence of twisted matter, $\phi_{A}$, the scalar potential was given in \cite{CLLSW} (see also \cite{LM}).  Here, we give the scalar potential when twisted matter is included, which is
\begin{equation}\label{chipot}
\begin{aligned}
	V_F=e^K & \Biggl\{ -3|W|^2 + |W-YW_s|^2 \\
	& + \sum\nolimits_A \widetilde{\Pi}_A^{-1} \frac{Y}{Y+p_A} \left|W_A  + \widetilde{\Pi}_A\bar{\phi}_A W + p_A\widetilde{\Pi}_A\bar{\phi}_A W_s \right|^2 \\
	& + \sum\nolimits_I\frac{Y}{Y+b_I+\sum\nolimits_A q_I^A X_A (Y+p_A)} \\
	&\qquad \times \biggl[ \left|x_IW_I+\sum\nolimits_A q_I^A\phi_AW_A - W - b_IW_s\right|^2 + x_I\sum\nolimits_A \Bigl|\bar{\phi}_{AI}W_I + W_{AI}\Bigr|^2\biggr]\Biggr\},
\end{aligned}
\end{equation}
where we have defined
\begin{equation}
	\widetilde{\Pi}_A\equiv\prod\nolimits_I x_I^{-q_I^A}.
\end{equation}


\subsection{The Linear Superfield Formalism} \label{lsf}
In the linear superfield formalism, in the form of the Bin\'etruy-Gaillard-Wu (BGW) model \cite{BGW}, the K\"ahler potential is \cite{GLM}
\begin{equation}\label{kpot}
	K=\ln(\ell)+g(\ell)-\sum\nolimits_I \ln x_I + \sum\nolimits_A X_A,
\end{equation}
where $\ell$ is the dilaton in the linear superfield formalism (its relation to the dilaton in the chiral superfield formalism is given in section \ref{?}) and  $g(\ell)$ is a nonperturbative correction that can stabilize the dilaton, which will be discussed in section \ref{dilstab}.  As we will see, the prescription for computing the scalar potential in this formalism requires temporarily replacing the dilaton with the (anti)chiral superfields $s$, $\bar{s}$.  Hence, the field dependence of the superpotential is
\begin{equation}
	W=W(s,t_I, \phi_{AI},\phi_A).
\end{equation}
Its general form will be given in section \ref{sp}.

The scalar potential may be derived through the following prescription.\footnote{Additional details may be found in Appendix A of \cite{BGN}.}  First form
\begin{equation}
	K^{(s)}\equiv k(s,\bar{s})-\sum\nolimits_I \ln x_I + \sum\nolimits_A X_A,
\end{equation}
which is identical to (\ref{kpot}) except that the $\ell$ dependence has been replaced with a dependence on the (anti)chiral superfields $s$, $\bar{s}$.  Then define the effective K\"ahler metric,
\begin{equation}\label{ekm}
	\widehat{K}_{m\bar{n}}\equiv K^{(s)}_{m\bar{n}} + \ell V^\tn{GS}_{m\bar{n}}, 
\end{equation}
whose inverse is $\widehat{K}^{m\bar{n}}$ and where $V^\tn{GS}$ was defined in (\ref{vgs}).  Both the effective K\"ahler metric and its inverse are given in appendix \ref{kml}.  It is conventional in this formalism to write the Green-Schwarz coefficient, $b_I$, as independent of $I$,\footnote{See foonote \ref{fngs}.}
\begin{equation}
	b_I= b.
\end{equation}
Standard compactifications lead to $b\leq30/8\pi^2$ \cite{gs}.  Again, the scalar potential is made up of the $F$-term and the $D$-term, but now the $F$-term is given by
\begin{equation}
	V_F=e^K\left[\widehat{K}^{m\bar{n}} (W_m+K^{(s)}_m W)(\overline{W}_{\bar{n}}+K^{(s)}_{\bar{n}} \overline{W})-3|W|^2\right],
\end{equation}
along with the replacements
\begin{equation}\label{repl}
	K^{(s)}_s\rightarrow-\ell, \qquad \widehat{K}_{s\bar{s}} \rightarrow \frac{\ell}{\partial K/\partial\ell}=\frac{\ell^2}{1+\ell g'(\ell)},
\end{equation}
where the prime denotes differentiation with respect to $\ell$.  One finds
\begin{equation}\label{pot}
\begin{aligned}
	V= e^K &\Biggl\{- 3|W|^2 + (\ell g' + 1)|W-\ell^{-1}W_s|^2 \\
	& + \sum\nolimits_A \widetilde{\Pi}^{-1}_A \frac{1}{1+p_A \ell}|W_A+\widetilde{\Pi}_A\bar{\phi}_A W +p_A\widetilde{\Pi}_A\bar{\phi}_A W_s|^2\\
	&+\sum\nolimits_I \frac{1}{1+b\ell+\sum\nolimits_B(1+p_B\ell)q_I^B X_B} \\
	&\qquad \times \biggl[ \Bigl| x_I W_I + \sum\nolimits_A q_I^A\phi_A W_A - W -bW_s \Bigr|^2+ x_I \sum\nolimits_A\Bigl| \bar{\phi}_{AI}W_I + W_{AI}\Bigr| ^2 \biggr] \Biggr\},
\end{aligned}
\end{equation}
where, as before, $\widetilde{\Pi}_A\equiv\prod\nolimits_I x_I^{-q_I^A}$.


\subsection{The Superpotential} \label{sp}

The superpotential is made up of both a perturbative term and a nonperturbative term,
\begin{equation}
	W=w_\tn{p} + w_\tn{np}.
\end{equation}
This terminology is meant to indicate that the perturbative term leads to (pertubative) loop corrections while the nonperturbative term is induced nonperturbatively.  

The perturbative term is essentially an arbitrary polynomial in the matter fields.  Its precise form is dictated by the requirement of (spacetime) modular invariance, to be discussed in section \ref{modin}.  If we denote both untwisted and twisted matter by $\phi_\alpha$, so that $\alpha=AI$, $A$, then, for either formalism, it is given by
\begin{equation}\label{gensp}
	w_\tn{p}=\sum\nolimits_m \lambda_m \left[\prod\nolimits_I \eta(t_I)^{-2}\right] \prod\nolimits_\alpha \phi_\alpha ^{n_m^\alpha} \prod\nolimits_J \eta(t_J)^{2 n_m^\alpha q_J^\alpha},
\end{equation}
where the $\lambda_m$ are constants, the $n_m^\alpha$ are nonnegative integers and $\eta(t_I)$ is the Dedekind eta function,
\begin{equation}
	\eta(t_I)=e^{-\pi t_I/12}\prod_{n=1}^\infty\left(1-e^{-2\pi nt_I}\right).
\end{equation}

The nonperturbative term follows from gaugino condensation \cite{gaugecon}.  A strongly coupled hidden sector gauge group, $G_a$, is expected to condense, analogously to QCD.  The corresponding gaugino condensates, $u_a=\langle \lambda_a^\alpha \lambda^a_\alpha \rangle$, pick up nonzero VEVs, spontaneously breaking supersymmetry.  If heavy enough they may be integrated out, inducing a nonperturbative contribution to the superpotential.   In the chiral superfield formalism this term takes the form \cite{BGW, BGN}
\begin{equation} \label{chiwnp}
	w_\tn{np}=-\frac{1}{4}e^{-K/2}\sum\nolimits_a b_a u_a, \qquad u_a=c e^{K/2}e^{-s/b_a} \prod\nolimits_I \left[\eta(t_I)\right]^{2(b_I-b_a)/b_a},
\end{equation}
where $b_a$ is the $\beta$-function coefficient of the condensing gauge group and $c$ is a constant.  By squaring $u_a$ we may write it in terms of $Y$ instead of $s$,
\begin{equation}
	|u_a|^2 = |c|^2 e^K e^{-Y/b_a} e^{-V^\tn{GS}/b_a} \prod\nolimits_I \left|\eta(t_I)\right|^{4(b_I-b_a)/b_a}.
\end{equation}
In the linear superfield formalism we have instead \cite{BGW, BGN}
\begin{equation} \label{linwnp}
	w_\tn{np}=-\frac{1}{4}e^{-K/2}\sum\nolimits_a b_a u_a, \qquad |u_a|^2 = |c|^2 e^K e^{-(1+f)/b_a\ell} e^{-V^\tn{GS}/b_a} \prod\nolimits_I \left|\eta(t_I)\right|^{4(b-b_a)/b_a},
\end{equation}
along with the replacement
\begin{equation}
	W_s \rightarrow \frac{1}{4}e^{-K/2} \sum\nolimits_a u_a.
\end{equation}
In (\ref{linwnp}) $f=f(\ell)$ is uniquely determined from the nonperturbative correction $g(\ell)$ though the differential equation and boundary conditions \cite{BGW}
\begin{equation}\label{f}
	\ell g' = f - \ell f', \qquad g(\ell=0)=f(\ell=0)=0,
\end{equation}
where a prime denotes differentiation with respect to $\ell$.


\subsection{The Chiral or Linear Superfield Formalism?}\label{?}
The dilaton in the two formalisms are related by \cite{BGW}
\begin{equation} \label{dilrel}
	\frac{\ell}{1+f}=\frac{1}{s+\bar{s}-V^\tn{GS}}=\frac{1}{Y},
\end{equation}
with $f$ defined in (\ref{f}).  If we ignore the nonperturbative term in the superpotential, then for identical perturbative terms the two formalisms are equivalent.  There exists a duality transformation, made up of (\ref{repl}) and (\ref{dilrel}), linking them, known as \textit{chiral linear duality} \cite{siegel, bgg}.  This is manifest in the classical limit, i.e. for $f=g=0$, but also holds true in the general case.  When the nonperturbative terms in the superpotential are included, it is unknown whether the two formalisms are equivalent \cite{bdqq}.  Even with such an equivalence, it may be simpler to build models in one formalism than in the other.  In the following subsections many of the results will be presented in both formalisms.\footnote{A major exception is section \ref{dilstab} where we consider only the linear superfield formalism.}  However, we will find, in many different cases, that inflation model building is simpler in the linear superfield formalism.  


\section{Canonical Normalization of the Inflaton} \label{norm}
Supergravity theories derived from string theory lead to noncanonically normalized kinetic terms.  To properly analyze inflation the canonical normalization of the inflaton must be determined.  Since the method we will use to build inflation models, to be described in section \ref{etaprob}, allows only K\"ahler moduli and untwisted matter (or some mixture thereof) to be the inflaton, we only consider the canonical normalization of these two types of fields.\footnote{In particular, we consider one of these fields (and not a linear combination of them) to be the inflaton, and the other to be stabilized.  Interesting alternatives can be found in, for example, \cite{CLLSW, ellis}.}

The kinetic terms will be given to lowest order in the Green-Schwarz coefficients.  This is largely unnecessary in the linear superfield formalism since it only requires dropping factors of $1+b\ell$, factors which have little effect on determining the canonically normalized field.  In the chiral superfield formalism, however, we must drop factors of $(Y+b_I)/Y$.  Dropping such factors make determining the canonically normalized field much easier, but make determining flat directions, as we will see in section \ref{etaprob}, difficult.

It is usually assumed that matter fields, both untwisted and twisted, have negligibly small values.  We cannot necessarily make this assumption for an untwisted matter field when it is the inflaton.  We will, however, always make this assumption for twisted matter.  Then, with the K\"ahler potential (\ref{chikpot}) or (\ref{kpot}), to lowest order in the Green-Schwarz coefficients, the kinetic terms for the untwisted matter fields, $\phi_{AI}$, are
\begin{equation} \label{kinterms}
	{\cal L}_\tn{kin} \supset \sum\limits_{A,I} \left( \frac{t_I+\bar{t}_I}{x_I^2}\partial_\mu \phi_{AI} \partial^\mu \bar{\phi}_{AI} + \frac{\phi_{AI}}{x_I^2} \partial_\mu t_I \partial^\mu \bar{\phi}_{AI}  + \frac{\bar{\phi}_{AI}}{x_I^2} \partial_\mu \phi_{AI} \partial^\mu \bar{t}_I \right).
\end{equation}
We assume the moduli fields, $t_I$, are stabilized during inflation (in section \ref{modin} we will see that they are usually stabilized at $O(1)$ values).  This means that they are effectively constant, their derivatives vanishing, allowing us to ignore the final two terms in (\ref{kinterms}).  We also assume that only the inflaton is comparable in size to the moduli fields, the rest of the untwisted matter fields being much smaller and negligible.  Without loss of generality, take the inflaton to be the $\phi_{11}$ field, whose kinetic term is then
\begin{equation}
	\frac{t_1+\bar{t}_1}{\left(t_1+\bar{t}_1-|\phi_{11}|^2\right)^2} \partial_\mu \phi_{11}\partial^\mu \bar{\phi}_{11}.
\end{equation}
If we ignore the phase, then the canonically normalized inflaton, $\phi$, is given by
\begin{equation}\label{can}
	|\phi_{11}|=\sqrt{t_1+\bar{t}_1}\tanh(\phi/\sqrt{2}),
\end{equation}
where we have used the assumption that the moduli fields are constant.

In the case where instead a K\"ahler moduli field is the inflaton we ignore both twisted and untwisted matter by assuming that they are small, and find, to lowest order in the Green-Schwarz coefficients, for the kinetic term
\begin{equation} \label{kintermsmod}
	{\cal L}_\tn{kin} \supset \sum\nolimits_I\frac{1}{(t_I+\bar{t}_I)^2}\partial_\mu t_I\partial^\mu \bar{t}_I.
\end{equation}
If we take the real part of $t_1$ to be the inflaton then the canonically normalized inflaton, $\phi$, is given by
\begin{equation}
	\tn{Re}(t_1)=e^{\sqrt{2}\phi}.
\end{equation}


\section{String Theory Requirements} \label{stringcon}
String theory places a number of requirements on its effective theory.  We list four that we will require inflation models to abide by.


\subsection{Modular Invariance} \label{modin}
If we denote both untwisted and twisted matter by $\phi_\alpha$, so that $\alpha=AI$, $A$, then modular transformations\footnote{By modular transformations we mean specifically spacetime T-duality transformations.} of the K\"ahler moduli and matter fields are defined by \cite{bl}
\begin{equation} \label{modtransf}
	t_I\rightarrow\frac{a_It_I-ib_I}{ic_It_I+d_I}, \qquad \phi_\alpha \rightarrow \phi_\alpha \prod\nolimits_I(ic_It_I+d_I)^{-q_I^\alpha},
\end{equation}
where
\begin{equation}
	a_Id_I-b_Ic_I=1,\qquad a_I,b_I,c_I,d_I\in\mathbb{Z}.
\end{equation}
Recall that the $q_I^\alpha$ are the modular weights and that for untwisted matter fields, $q_J^\alpha=q_J^{AI}=\delta_J^I$, while for twisted matter fields, $q_I^\alpha=q_I^A\geq0$.\footnote{See footnote \ref{fnmw}.}  The K\"ahler potential and superpotential also undergo modular transformations, which are a special case of a K\"ahler-Weyl transformation \cite{zumino, WB},
\begin{equation}
	K\rightarrow K+\sum\nolimits_I\ln|ic_It_I+d_I|^2, \qquad W\rightarrow W \prod\nolimits_I(ic_It_I+d_I)^{-1}.
\end{equation}
Thus, the superpotential transforms with modular weight equal to 1.  An important function with modular weight equal to $-1/2$ is the Dedekind eta function,
\begin{equation}
	\eta(t_I)=e^{-\pi t_I/12}\prod_{n=1}^\infty\left(1-e^{-2\pi nt_I}\right), \qquad \eta(t_I) \rightarrow \eta(t_I) (ic_It_I+d_I)^{1/2},
\end{equation}
which tells us that the superpotential transforms as $\eta(t_I)^{-2}$ and matter fields transform as $\eta(t_I)^{-2q_I^\alpha}$.

Heterotic string theory is known to be modular invariant to all orders in perturbation theory \cite{gmr}.  This means that the effective theory must be as well.  This is assured at tree level since modular transformations are just special cases of K\"ahler-Weyl transformations and K\"ahler-Weyl transformations are always symmetries of a tree level supergravity Lagrangian \cite{zumino, WB}.  Field theory loop corrections in general break the modular symmetry, leading to a modular anomaly.  This anomaly is partially\footnote{Threshold corrections from integrating out heavy fields cancel the remainder of the anomaly \cite{dkl, bl}.} canceled by introducing the Green-Schawz counterterm (\ref{vgs}) \cite{gs}.  

We saw in section \ref{scalarpots} that the Green-Schwarz counterterm is introduced differently in the chiral and linear superfield formalisms.  In the chiral superfield formalism, the dilaton, $s+\bar{s}$, is modular invariant at tree level, but not so at loop level.  For this reason, the Green-Schwarz counterterm was introduced so that the combination in (\ref{chiy}) is modular invariant and the K\"ahler potential (\ref{chikpot}) transforms correctly.  An important advantage of the linear superfield formalism is that the dilaton, $\ell$, is modular invariant to all orders in perturbation theory.  This allows the Green-Schwarz counterterm to be introduced as in (\ref{ekm}).  As we will see, this will lead to simplifications.

Modular invariance also dictates the form of the perturbative term in the superpotential \cite{BGW, GLM}.  We wrote down its form in (\ref{gensp}).  The eta functions in front make sure the superpotential transforms with modular weight 1, while the eta functions on the end cancel the modular transformations of the $\phi_\alpha$'s.


\subsection{Absence of Mass Terms} \label{mterms}

Massive states in string theory have masses on the order of the string scale ($\sim\! 10^{17}$ GeV).  Since an effective theory is only relevant far below this scale, all massive fields must be integrated out (leading to threshold corrections).  Hence, the effective theory contains only massless fields (at least, before any fields pick up nonzero VEVs) and cannot contain mass terms.  From (\ref{chipot}) or (\ref{pot}) we see that the superpotential and its derivatives enter the scalar potential squared.  For there to be no dimension two terms in the scalar potential, i.e. no mass terms, each term in the superpotential must be dimension three or greater.  Note that this applies only to matter fields, $\phi_\alpha$, with moduli not included in the counting.


\subsection{Dilaton Stabilization} \label{dilstab}

The method we will use for building inflation models, to be described in section \ref{etaprob}, does not allow the dilaton to be the inflaton.\footnote{In general, the dilaton as the inflaton is problematic \cite{BS}.}  This means that the dilaton must be stabilized, otherwise it can destroy inflation \cite{BS}.  By stabilized we mean that during inflation the dilaton potential must contain a stable minimum.  The method we use to achieve this, known as \textit{K\"ahler stabilization} \cite{BGW, Casas, Shenker, Silverstein}, has been worked out in the linear superfield formalism in some detail \cite{BGW, CG}, where it is simpler.  It has also been considered in a chiral superfield formalism without modular invariance \cite{Casas}.  We consider only the linear superfield formalism.

We make two further requirements.  First, in the true vacuum the dilaton potential must have a stable minimum with vanishing vacuum energy,\footnote{A related issue concerning dilaton stabilization in the BGW model was considered in \cite{skinner}.} and second, the coupling constant for the gauge fields at the string scale, $g_s$, must take the supersymmetric GUT value, $\approx 0.7$.

In the models we will consider the dilaton dependence during inflation is contained in the overall factor
\begin{equation} \label{dilpot}
	V^{(\ell\:\!)}_\tn{inf} = \frac{e^{g}\ell}{1+b\ell},
\end{equation}
where $g=g(\ell)$ is the nonperturbative correction in (\ref{kpot}).  In the true vacuum, using gaugino condensates to break supersymmetry and making the usual assumption that all matter fields vanish, we find that the K\"ahler moduli are stabilized at the fixed point $t_I=1$ \cite{BGW}.  The scalar potential in the true vacuum is then (see, for example, \cite{CG})
\begin{equation} \label{vacpot}
	V_0\propto\frac{1}{b_a^2\ell^2}(f-f'\ell+1)(1+b_a\ell)^2-3,
\end{equation}
where $b_a$ is the $\beta$-function coefficient for the hidden sector condensing gauge group which produces the gaugino condensates, with phenomenologically preferred values $0.03\apprle b_a \apprle 0.04$ \cite{bn}, and $f=f(\ell)$ was defined in (\ref{f}).  Finally, the coupling constant for the gauge fields at the string scale, $g_s$, is given by \cite{BGW}
\begin{equation}\label{gs}
	g_s^2=\frac{2\ell}{1+f},
\end{equation}
where the right hand side is to be evaluated at the true vacuum.

\begin{figure}
	\centering
		\includegraphics[width=6in,viewport=30 1 720 156,clip]{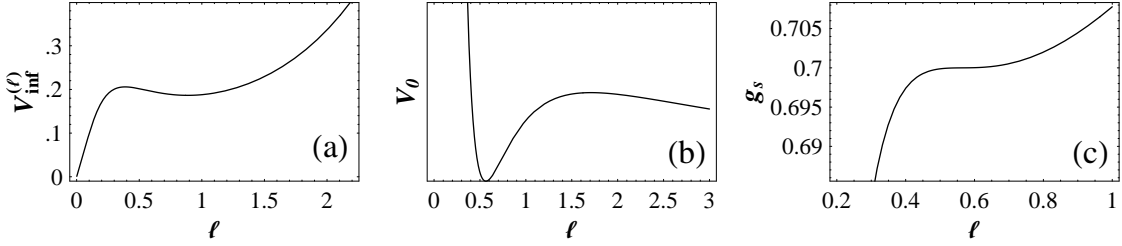}
	\caption{(a) The dilaton potential during inflation (\ref{dilpot}) with a stable minimum at $\langle \ell \rangle_\tn{inf}=0.87$ .  (b) The dilaton potential in the true vacuum (\ref{vacpot}) with a stable minimum and vanishing vacuum energy at $\langle\ell\rangle_0=0.56$.  (c) Evaluated at $\langle\ell\rangle_0=0.56$, the coupling constant at the string scale (\ref{gs}) is $g_s=0.7$.}
	\label{plotdil1}
\end{figure}

A similar (and more comprehensive) analysis of these requirements for the dilaton was made in \cite{CG}, where additional phenomenological constraints were mentioned.  Following \cite{GLM, CG} we use only the two leading terms for the nonperturbative parameters \cite{Shenker},
\begin{equation}\label{arbf}
	f(\ell)=B\left(1+A\frac{1}{\sqrt{b\ell}}\right)e^{-1/\sqrt{b\ell}}.
\end{equation}
In figure \ref{plotdil1}
 we have plotted (\ref{dilpot}), (\ref{vacpot}) and (\ref{gs}) for the values
\begin{equation}\label{npval}
	A=-0.27, \qquad B=26.8.	
\end{equation}
Figure \ref{plotdil1}(a) shows a stable minimum during inflation at $\langle \ell \rangle_\tn{inf}=0.87$, figure \ref{plotdil1}(b) shows a stable minimum with vanishing vacuum energy in the true vacuum at $\langle\ell\rangle_0=0.56$, and figure \ref{plotdil1}(c) shows, when evaluated at $\langle\ell\rangle_0=0.56$, that the coupling constant at the string scale is $0.7$.


\subsection{K\"ahler Moduli Stabilization} \label{modstab}

In addition to the dilaton, the three K\"ahler moduli, $t_I$, as long as they are not the inflaton, must be stabilized.  In modular invariant theories we are assured that the scalar potential will have stationary points at $t_I=1$, $\exp(i\pi/6)$.  These two points correspond to fixed points of the modular transformation (\ref{modtransf}).  The important question is whether the stationary points are (stable) minima.

\begin{figure}
	\centering
		\includegraphics[width=5.75in,viewport=33 1 715 218,clip]{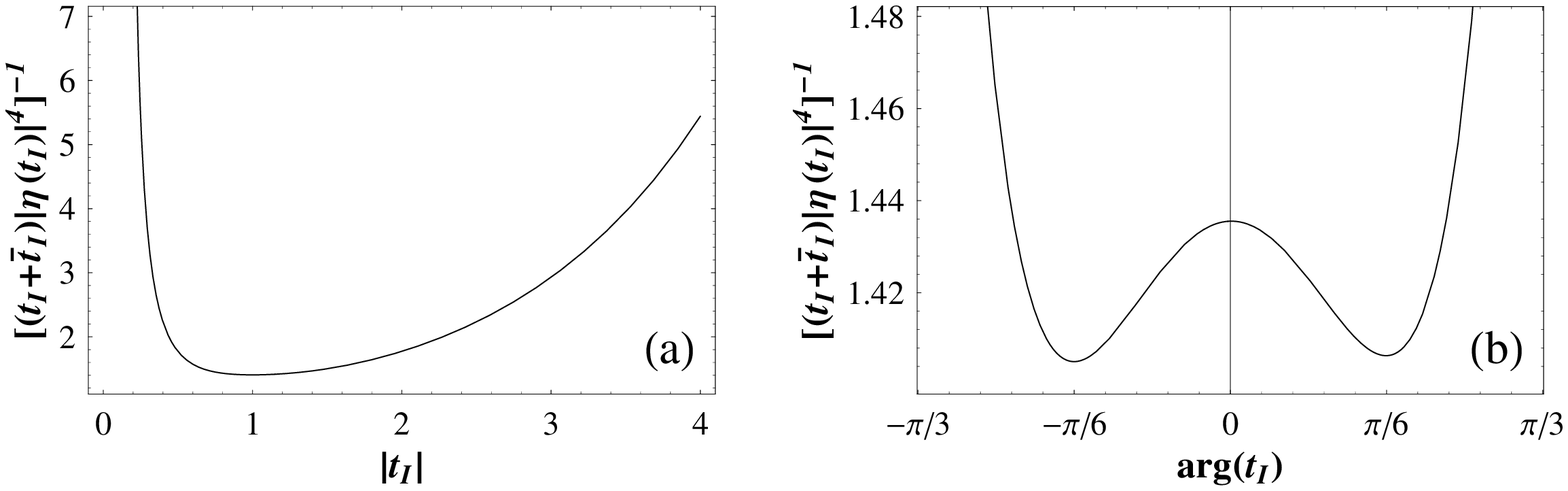}
	\caption{(a) A plot of (\ref{moddep}) with $\arg(t_I)=\pi/6$.  (b) A plot of (\ref{moddep}) with $|t_I|$=1.  A stable minimum is clearly seen at $t_I=\exp(i\pi/6)$.}
	\label{plotded1}
\end{figure}

In the models we will consider the K\"ahler moduli dependence is often of the form
\begin{equation}\label{moddep}
	\left[x_I|\eta(t_I)|^4\right]^{-1}=\left[(t_I+\bar{t}_I)|\eta(t_I)|^4\right]^{-1},
\end{equation}
where we have made the usual assumption that the untwisted matter fields, $\phi_{AI}$, are negligible compared to the moduli.  Setting $t_I=\exp(i\pi/6)+\sqrt{3}\delta$ one finds \cite{CLLSW}
\begin{equation}\label{modstabeq}	\left[(t_I+\bar{t}_I)\left|\eta(t_I)\right|^4\right]^{-1}=\frac{1}{\sqrt{3}}\left|\eta\left(e^{i\pi/6}\right)\right|^{-4}\left[1+|\delta|^2+O(\delta^3)\right],
\end{equation}
showing that $t_I=\exp(i\pi/6)$ corresponds to a stable minimum ($t_I=1$ corresponds to a saddle point).  Alternatively this can be shown by plotting (\ref{moddep}).  Figure \ref{plotded1}(a)
 is a plot of (\ref{moddep}) with $\arg(t_I)=\pi/6$ and figure \ref{plotded1}(b) is a plot of (\ref{moddep}) with $|t_I|=1$.  It is clear that $t_I=\exp(i\pi/6)$ corresponds to a stable minimum.  


\section{Generating VEVs} \label{vevs}
In section \ref{mterms} we required each term of the superpotential to have three or more matter fields.  Only one of these fields will be the inflaton.  The remainder will be given VEVs.  We consider two methods for generating VEVs which we present in the linear superfield formalism.  In appendix \ref{chivevs} we reproduce them in the chiral superfield formalism.


\subsection{$D$-term VEVs} \label{dvevs}
In many orbifold compactifications there is an anomalous $U(1)$ gauge group \cite{giedt}.  Canceling the anomaly requires a Green-Schwarz counterterm which leads to a Fayet-Illiopoulos contribution to the $D$-term \cite{dsw}.  A natural way for fields to obtain VEVs is by having them cancel such a $D$-term.  The $D$-term contribution to the scalar potential is
\begin{equation} \label{dpotgen}
	V_D=\frac{1}{2}g^2\left(\sum\nolimits_\alpha q_\alpha K_\alpha \phi_\alpha + \xi_D\right)^2,
\end{equation}
where $q_\alpha$ is the $U(1)$ charge (and should not be confused with the modular weight) and, in the linear superfield formalism, the Fayet-Illiopoulos term is
\begin{equation} \label{FI}
	\xi_D=\frac{2\ell \tn{Tr}(Q)}{192\pi^2},
\end{equation}
where $\tn{Tr}(Q)=\sum\nolimits_\alpha q_\alpha \sim 100$ \cite{giedt}.  With the K\"ahler potential (\ref{kpot}), the $D$-term (\ref{dpotgen}) becomes
\begin{equation}\label{dpot}
	V_D=\frac{1}{2}g^2\left[\sum\nolimits_\alpha\left(\prod\nolimits_I x_I^{-q_I^\alpha}\right)q_\alpha |\phi_\alpha|^2 + \xi_D\right]^2. 
\end{equation}
To avoid $D$-term supersymmetry breaking during inflation the matter fields must pick up the modular invariant VEVs,
\begin{equation}\label{dvev}
	|\langle\phi_\alpha\rangle|^2=f_\alpha \ell \prod\nolimits_I x_I^{q_I^\alpha},
\end{equation}
where $f_\alpha$ is a constant, to cancel (\ref{dpot}).


\subsection{$F$-term VEVs} \label{fvevs}
It is also possible to induce VEVs using the $F$-term of the scalar potential \cite{GLM}.  Consider three fields, $\chi$, $\phi_2$, $\phi_3$.  Assume $\phi_2$, $\phi_3$ pick up nonzero VEVs: $\langle\phi_{2}\rangle$, $\langle\phi_{3}\rangle \neq 0$ (for example, by canceling a $D$-term).  Now form the modular invariant expression
\begin{equation} \label{gammagen}
	\Gamma=\chi\phi_2\phi_3 \prod\nolimits_I\eta(t_I)^{2\sum\nolimits_\beta q_I^\beta}, \qquad \beta=\chi,2,3,
\end{equation}
to be used in the superpotential
\begin{equation}\label{wvevsp}
	W(\Gamma)= \left[\psi\phi_2\phi_3\prod\nolimits_I\eta(t_I)^{-2(1-\sum_\gamma q_I^\gamma)}\right]\sum_{n=0} c_n \Gamma^n, \quad \gamma = \psi, 2, 3,
\end{equation}
where the $c_n$ are constants.  It can be shown that, upon plugging this superpotential into the scalar potential, $\langle \psi \rangle =0$, and thus
\begin{equation} \label{wwppot1}
	V=e^K \prod\nolimits_J x_J^{q_J^\psi}\left|\phi_2\phi_3\prod\nolimits_I\eta(t_I)^{-2(1-\sum_\gamma q_I^\gamma)}\left[c_0+\sum_{n=1}c_n \Gamma^n\right]\right|^2.
\end{equation}
Since $\phi_2$, $\phi_3$ are assumed to have nonzero VEVs, $\langle \chi \rangle$ is determined by
\begin{equation} \label{vaccon}
	c_0+\sum_{n=1}c_n \Gamma^n=0.
\end{equation}
The only way that (\ref{vaccon}) could be satisfied is if \cite{GLM}
\begin{equation}
	\Gamma = f^3,
\end{equation}
where $f$ is a constant, and thus
\begin{equation} \label{phivevgen}
	|\langle\chi\rangle|^2 = |f|^6 \left|\left< \phi_2 \phi_3 \prod\nolimits_I\eta(t_I)^{2\sum\nolimits_\beta q_I^\beta} \right> \right|^{-2}.
\end{equation}


\section{Inflation Model Building} \label{modelbuilding}

In this section we consider possible directions for inflation model building in the supergravity theory described above.

For convenience we will limit ourselves to untwisted matter and will suppress the $A$ index.  For example, $\phi_1$ corresponds to an untwisted matter field from the $I=1$ moduli sector.  We will also make use of the shorthand $\eta_I=\eta(t_I)$ for the Dedekind eta function.


\subsection{The $\eta$-problem and a Method for Building Inflation Models} \label{etaprob}

As a first step in $F$-term inflation model building in supergravity, one must overcome the $\eta$-problem, which may be understood as follows.  With the help of a K\"ahler-Weyl transformation and a holomorphic field redefinition a K\"ahler potential may be written as
\begin{equation}
	K=\sum\nolimits_a |\phi_a|^2 + \cdots,
\end{equation}
where dots represent additional and, for the purpose of this subsubsection, irrelevant terms.  The $F$-term contribution to the scalar potential, as we have seen above, is of the form
\begin{equation}
	V_F=e^K ( \cdots ).
\end{equation}
The slow roll parameter $\eta$, defined in (\ref{srpar}), is then given by
\begin{equation} \label{etaprobeta}
	\eta=1+\cdots.
\end{equation}
Recall that slow roll inflation requires $|\eta| \ll 1$, and therefore, barring model dependent cancellations of the 1 in (\ref{etaprobeta}), slow roll inflation will not occur in a generic supergravity scalar potential.  This is the $\eta$-problem.

The method we use to solve the $\eta$-problem was proposed in \cite{CLLSW, Stewart}, and is particularly suited for orbifold compactifications of string theory.  Consider three fields, $\phi$, $\psi$, $\chi$, where $\phi$ is the inflaton.  Take $\psi$ to have a small, perhaps vanishing, VEV during inflation, designated by
\begin{equation}
	\langle \psi \rangle \sim 0.
\end{equation}
By small we mean that $\psi$ is completely negligible in the scalar potential so that any term containing $\psi$ can be ignored (i.e. set to zero).  We take this also to mean that any function which contains $\psi$ can also be ignored.  In particular, the superpotential, $W(\phi,\psi,\chi)$, which is assumed to be a polynomial function of the fields in which every term contains $\psi$, during inflation satisfies
\begin{equation}
	\langle W(\phi,\psi,\chi) \rangle \sim 0.
\end{equation}
$\phi$ and $\chi$ derivatives do not effect the overall $\psi$ dependence of $W$ and so are small as well, $\langle W_\phi \rangle$, $\langle W_\chi \rangle \sim0$.  However, a $\psi$ derivative of $W$ removes a factor of $\psi$, and is therefore not small,
\begin{equation}
	\left<W_\psi \right> \not\sim 0.
\end{equation}
As shown by Stewart \cite{Stewart}, for a K\"ahler potential of the form\footnote{Stewart actually gives the general form that the K\"ahler potential may take, which includes (\ref{kpotgen}) \cite{Stewart}.} (\ref{kpotgen}), a $W_\psi$ which leads to inflation in \textit{global} supersymmetry will also lead to inflation in supergravity where the inflaton can be a K\"ahler moduli or an untwisted matter field.  Thus, inflation model building in supergravity has been rendered equivalent to inflation model building in global supersymmetry, where there is no $\eta$-problem.


\subsection{Preserving Flat Directions I} \label{flatI}

As an example of what these assumptions can do, we review the superpotential presented in \cite{GLM},
\begin{equation} \label{flatsp}
	W =\lambda \psi_{1}\eta_2^{-2}\eta_3^{-2},
\end{equation}
which is of the form (\ref{gensp}) and where $\lambda$ is a constant.  This superpotential does not satisfy the requirements of section \ref{mterms}. It will be generalized in section \ref{dinvevs} so that it does.  For now, we consider it for illustrative purposes.  Plugging the superpotential (\ref{flatsp}) into (\ref{pot}) we obtain the scalar potential in the linear superfield formalism.  It was not shown in \cite{GLM} under what conditions $| \psi_1 | = 0$ will correspond to a stable minimum so that the framework of the previous subsection may be used. In appendix \ref{apenetaprob} we show explicitly the conditions required.  Assuming these conditions, so that we may set $| \psi_1 | = 0$, the scalar potential collapses down to \cite{GLM}
\begin{equation}
	V=e^K x_1 \frac{|\lambda|^2}{| \eta_2\eta_3|^4}.
\end{equation}
For the K\"ahler potential (\ref{kpot}) this becomes \cite{GLM}
\begin{equation}
	V=\frac{\ell e^g}{1+b\ell}\frac{|\lambda|^2}{x_2 x_3|\eta_2\eta_3|^4 }. \label{flat}
\end{equation}
In the chiral superfield formalism we instead find
\begin{equation}
		V=\frac{1}{Y+b_1}\frac{|\lambda|^2}{x_2 x_3|\eta_2\eta_3|^4 }. \label{chiflat}
\end{equation}

Notice that (\ref{flat}) is completely independent of moduli and untwisted matter from the first moduli sector ($t_1$ and $\phi_{1A}$).  Any combination of these fields corresponds to perfect flat directions of the potential.  The framework of section \ref{etaprob} has allowed us to preserve flat directions, canceling their generic lifting by supergravity, thus solving the $\eta$-problem.

In (\ref{chiflat}) there is an explicit dependence on these fields through the Green-Schwarz counterterm (\ref{vgs}) contained in $Y$, which appears to lift the flat directions and contradict our statement in section \ref{?} that the two formalisms are equivalent.  In fact the flat directions do exist in (\ref{chiflat}); it is just more difficult to determine what they are.  To do so requires diagonalizing the K\"ahler metric, equivalent to canonically normalizing the fields, just as we did in section \ref{norm}, but now with the Green-Schwarz coefficients included---not an easy thing to do in the chiral superfield formalism.

Finally, we note two things.  First, the dilaton dependence of (\ref{flat}) is precisely what we showed in section \ref{dilstab} could stabilize the dilaton during inflation and in the true vacuum, and second, the K\"ahler moduli dependence of (\ref{flat}) is precisely what we showed in section \ref{modstab} would stabilize K\"ahler moduli.


\subsection{Preserving Flat Directions II} \label{dinvevs}

The superpotential in section \ref{flatI} may be generalized so that it satisfies the requirements of section \ref{mterms} \cite{GLM}.  To do so, consider the superpotential
\begin{equation} \label{wwpsp}
	W_0=w+w',
\end{equation}
where
\begin{equation}\label{wsp}
	w=\lambda \psi_1\chi_2\phi_2\phi_3\eta_2^2.
\end{equation}
Assume $\phi_2$, $\phi_3$ are charged under an anomalous $U(1)$ gauge group and pick up VEVs to cancel the $D$-term, as explained in section \ref{dvevs}.  If (\ref{wsp}) were the entire superpotential (so that we did not have the contribution $w'$ in (\ref{wwpsp})), we would find (under certain assumptions; see appendix \ref{apenetaprob}) that $\langle \psi_1 \rangle =0$, which would give
\begin{equation}
	V=e^K x_1 |\langle \chi_2\phi_2\phi_3 \rangle |^2 |\eta_2|^4.
\end{equation}
It would appear then that $\langle \chi_2 \rangle =0$, but we are going to use $w'$ in (\ref{wwpsp}) to induce a ($F$-term) VEV for $\chi_2$.  Following (\ref{wvevsp}) we take
\begin{equation} \label{wpsp}
	w'=\psi'_1\phi'_2\phi'_3 \sum_{n=0}c_n \left(\chi_2 \phi'_2 \phi'_3 \eta_2^4\eta_3^2\right)^n
\end{equation}
where the primed fields in (\ref{wpsp}) and the unprimed fields in (\ref{wsp}) are distinct, the $c_n$ are constants and $\phi'_2$, $\phi'_3$ are assumed to pick up $D$-term VEVs.  Plugging the complete superpotential (\ref{wwpsp}) (made up of (\ref{wsp}) and (\ref{wpsp})) into the scalar potential, we find that $\langle \psi_1 \rangle = \langle \psi'_1 \rangle =0$ (see appendix \ref{appdinvevs} for details), which gives
\begin{equation} \label{wwppot}
	V=e^K x_1\left[\Bigl| \lambda \chi_2\phi_2\phi_3 \eta_2^2\Bigr|^2 + \biggl|\phi'_2\phi'_3 \Bigr[c_0+\sum_{n=1}c_n \left(\chi_2 \phi'_2 \phi'_3 \eta_2^4\eta_3^2\right)^n\Bigr]\biggr|^2\right].
\end{equation}
Although $\langle \psi_1 \rangle$ cannot be determined until the $c_n$ are specified, we might imagine that $V$ is minimized when the second term in (\ref{wwppot}) vanishes \cite{GLM}.  This is plausible since the term lowest order in the fields is the $c_0$ term and we might expect that $\langle \psi_1 \rangle$ is determined by its cancellation.  This was the case considered in section (\ref{fvevs}) and we assume it to be true here \cite{GLM}.  From (\ref{phivevgen}) we then have \cite{GLM}
\begin{equation} \label{psivev}
	|\langle\chi_2\rangle|^2 = |f|^6 |\langle \phi'_2 \phi'_3 \eta_2^4\eta_3^2 \rangle |^{-2}.
\end{equation}
The potential is then
\begin{equation} \label{psipot}
	\langle V \rangle = |\lambda|^2|f|^6 \frac{f_2 f_3}{f'_2 f'_3} \frac{\ell e^g}{1+b\ell} \frac{1}{x_2 x_3 |\eta_2 \eta_3|^4},
\end{equation}
where we have plugged in the $D$-term VEVs of $\phi_2$, $\phi_3$, $\phi'_2$, $\phi'_3$ using (\ref{dvev}).


\subsection{Loop Corrections} \label{loop}

In the next subsection we will consider a model in which the inflaton enters at loop level.  In this subsection we review the relevant one loop corrections to the scalar potential.

In the full supergravity theory there are numerous terms in the one loop correction to the scalar potential \cite{looppapers1, looppapers2}.  Fortunately, for the models we consider, there will be only a single term that contributes to the inflaton mass \cite{GMO, looppapers1}.  To write down this term we begin by reviewing K\"ahler covariant notation (see, for example, \cite{looppapers1}).  First, we define
\begin{equation}
	A \equiv e^K W,
\end{equation}
with corresponding covariant derivatives
\begin{equation}
	A_m = D_m A = \partial_m A = \frac{\partial}{\partial\phi_m}A, \qquad A_{mn}=D_m D_n A = \partial_m A_n - \Gamma^\ell_{mn}A_\ell,
\end{equation}
where the connection is defined by
\begin{equation}
	\Gamma^\ell_{mn}=K^{\ell\bar{p}}\partial_m K_{n\bar{p}}.
\end{equation}
With this notation we can now compactly write down the relevant loop correction.  It is contained in the expression \cite{GMO, looppapers1}
\begin{equation} \label{loopcor}
	\Delta V= 2\langle V \rangle \varepsilon e^K A_{ij}\bar{A}^{ij}, \qquad \varepsilon=\frac{\ln \Lambda}{32\pi^2},
\end{equation}
where $\Lambda$ is the momentum cutoff, $\langle V \rangle$ is the tree level $F$-term contribution to the scalar potential and
\begin{equation}
	\bar{A}=e^K\overline{W}, \qquad \bar{A}^{mn}=K^{m\bar{\ell}}K^{n\bar{p}}\bar{A}_{\bar{\ell}\bar{p}}.
\end{equation}

Consider now a superpotential of the form
\begin{equation} \label{wloop}
	W=\lambda\phi_1\phi_2\phi_3 + \phi'_1(\cdots) + \phi''_1(\cdots),
\end{equation}
where $\phi_1$ is the inflaton, $\phi_2$, $\phi_3$, $\phi'_1$, $\phi_1''$ are zero during inflation and the dots are any set of fields not contained in the first term nor from the same moduli sector as the inflaton (in this case $I=1$).  If (\ref{wloop}) leads to the tree level scalar potential $\langle V \rangle$ then by including the loop correction (\ref{loopcor}) it can be shown that
\begin{equation}
	V=\langle V \rangle \left[1-4|\lambda|^2\varepsilon\frac{|\phi_1|^2}{t_1+\bar{t_1}-|\phi_1|^2}\right] = \langle V \rangle \Bigl[1-4|\lambda|^2\varepsilon \sinh^2(\phi)\Bigr],
\end{equation}
where $\phi$ is the canonically normalized inflaton.  Expanding this to quadratic order we have
\begin{equation}
	V= \langle V \rangle \bigl(1-4|\lambda|^2\varepsilon\phi^2\bigr),
\end{equation}
which is an example of a so called \textit{small field} model of inflation \cite{lr, MR}.


\subsection{Loop Potential} \label{loopmod1}

In the previous subsection we saw that to construct a model in which the inflaton dependence shows up at loop level there must be a nonzero tree level vacuum energy.  Obviously the inflaton must also be missing from the tree level potential.  Both of these requirements are satisfied by the potential in section \ref{dinvevs}.  In this subsection we combine the works of \cite{GLM, GMO} and consider the loop corrections to that potential.

The superpotential we use is
\begin{equation}
	W = \lambda'\phi_1\phi''_2\phi''_3 + W_0,
\end{equation}
where $W_0$ was given in (\ref{wwpsp}), $\phi_1$ is the inflaton and both $\phi''_2$ and $\phi''_3$ have vanishing VEVs.  This is of the form (\ref{wloop}) and it can be shown that with the fields at their VEVs the tree level potential comes out as if the superpotential were just $W_0$, which was given in (\ref{psipot}) (see appendix \ref{apploopmod1}),
\begin{equation}
		\langle V \rangle = |\lambda|^2|f|^6 \frac{f_2 f_3}{f'_2 f'_3} \ell e^g \frac{1}{x_2 x_3 |\eta_2 \eta_3|^4}.
\end{equation}
The inflaton dependence then enters at loop level, as explained in the previous subsection, yielding
\begin{equation}\label{looppot}
	V = \langle V \rangle \left[1-4|\lambda'|^2\varepsilon \sinh^2(\phi)\right].
\end{equation}
This potential stabilizes the dilaton, $t_1$, $t_2$, but does not induce a potential for $t_1$.  For small $\phi$ during inflation,
\begin{equation} \label{looppotsf}
	V = \langle V \rangle \left(1-4|\lambda'|^2\varepsilon \phi^2\right).
\end{equation}


\subsection{Phenomenology}

The inflaton potential in (\ref{looppotsf}) is of the form
\begin{equation}\label{sfim}
	V=V_0 \left[ 1 - \left( \frac{\phi}{\mu} \right)^2 \right],
\end{equation}
with
\begin{equation}
	V_0=|\lambda|^2|f|^6 \frac{f_2 f_3}{f'_2 f'_3} \ell e^g \frac{1}{x_2 x_3 |\eta_2 \eta_3|^4}, \qquad \mu=\frac{1}{\sqrt{4|\lambda|^2\varepsilon}}.
\end{equation}
For large $\mu$ (\ref{sfim}) is an example of a small field model of inflation \cite{lr, MR}.  This is precisely the situation we have since $\varepsilon$ is small, it being a loop factor.  During inflation $\phi/\mu$ is usually small enough that $V\approx V_0$.  The slow roll parameters (\ref{srpar}) are then given by
\begin{equation}
	\epsilon = \frac{2}{\mu^2} \left(\frac{\phi}{\mu}\right)^2, \qquad \eta = -\frac{2}{\mu^2}, \qquad \xi^2 = 0.
\end{equation}
For inflation to be possible we require
\begin{equation} \label{sfreq}
	\mu \gg \sqrt{2}
\end{equation}
so that $|\eta| \ll 1$.  Slow roll inflation ends at
\begin{equation} \label{sfphie}
	\phi_e = \frac{1}{\sqrt{2}}\mu^2,
\end{equation}
rolling from left to right.  In lieu of (\ref{sfreq}) inflation is ending at large field values.  If we wish to restrict ourselves to small field values we must imagine that some other mechanism is acting to end inflation earlier, such as the hybrid inflation mechanism \cite{linde, lr}.  We will therefore write the remaining equations in terms of an arbitrary ending field value, $\phi_e$.

The number of $e$-folds, $N$, from $\phi=\phi_*$ until the end of inflation at $\phi=\phi_e$ can be computed with (\ref{efoldeq}).  Inverting this result yields
\begin{equation}
	\phi_* = \phi_e e^{-2N/\mu^2}.
\end{equation}
With this we may write the spectral index, its running and the tensor fraction (\ref{infeqs}) as
\begin{align}
	n &= 1 - \frac{4}{\mu^2}-\frac{12}{\mu^2}\left(\frac{\phi_e}{\mu}\right)^2e^{-4N/\mu^2}, \\
	\alpha &= -\frac{96}{\mu^4}\left(\frac{\phi_e}{\mu}\right)^4 e^{-8N/\mu^2} - \frac{64}{\mu^4}\left(\frac{\phi_e}{\mu}\right)^2 e^{-4N/\mu^2}, \\
	r &= \frac{32}{\mu^2} \left(\frac{\phi_e}{\mu}\right)^2 e^{-4N/\mu^2}.
\end{align}
The COBE normalization (\ref{cobe}) depends on the value of $\mu$.  For $\mu\sim10$ we have roughly
\begin{equation}
	V_0 \sim |f|^6|\lambda'|^2 \sim 10^{16}\tn{ GeV}.
\end{equation}


\section{Conclusion}\label{con}
We have described methods for building ``semi-realistic" models of $F$-term inflation.  By semi-realistic we mean that they are built in, and obey the requirements of, ``semi-realistic" particle physics models, taken here to be effective supergravity theories derived from orbifold compactifications of the weakly coupled heterotic string.  We reviewed those aspects of the supergravity theories relevant for inflation model building in both the chiral and linear superfield formalisms. This included scalar potentials with a comprehensive matter content, string theory requirements that the effective supergravity theories should obey and various tools and methods for building inflation models.  

In the course of this review we found that inflation model building is much simpler in the linear superfield formalism.  In particular, canonical normalization of the fields, determination of flat directions, moduli stabilization and generation of VEVs was found to be simpler.  The reason for this is the manner in which the linear superfield formalism includes the Green-Schwarz counterterm.

After reviewing previous work on inflation model building in these supergravity theories we combined them to construct a small field model of inflation in which the inflaton enters at loop order.  This model is incomplete in that it does not have a natural end to inflation, such as through a hybrid mechanism.  Regardless, it is illustrative of directions one may take in building inflation models.  Building such models is not simple and more will have to be done to build more realistic models, but we hope that we have been able to offer methods for how inflation model building can take into account details of a particular underlying particle physics model.

\section*{Acknowledgments}
I am grateful to Mary K. Gaillard for her help while preparing this paper.  This work was supported in part by the Director, Office of Science, Office of High Energy and Nuclear Physics, Division of High Energy Physics of the U.S. Department of Energy under Contract DE-AC02-05CH11231, in part by the National Science Foundation under grant PHY-0098840.

\appendix

\section{K\"ahler Metrics}

The K\"ahler metric, $K_{m\bar{n}}$, is defined by
\begin{equation}
	K_{m\bar{n}} = \frac{\partial}{\partial \phi_m} \frac{\partial}{\partial \bar{\phi}_{\bar{n}}} K,
\end{equation}
where $K$ is the K\"ahler potential.  $K^{m\bar{n}}$ is the inverse K\"ahler metric, defined by
\begin{equation}
	K_{m\bar{p}}K^{\bar{p}n}=\delta_m^n, \qquad \left(K^{m\bar{n}} \right)^* = K^{\bar{m}n},
\end{equation}
with the star denoting complex conjugation.


\subsection{The Chiral Superfield Formalism} \label{kmc}

The K\"ahler potential in section \ref{csf} is
\begin{equation}
	K = -\ln Y - \sum\nolimits_I \ln x_I + \sum\nolimits_A X_A.
\end{equation}
We define the following quantities,
\begin{align}
	\widetilde{b}_I &\equiv 1 + Y^{-1}b_I \\
	\widetilde{P}_A &\equiv 1 + Y^{-1}P_A \\
	\alpha_I &\equiv b_I + \sum\nolimits_A q_I^A P_A  X_A \\
	\beta_I &\equiv \widetilde{b}_I + \sum\nolimits_D q_I^D \widetilde{P}_D X_D \\
	\widetilde{\Pi}_A &\equiv \Pi_I x_I^{-q_I^A}.
\end{align}
The K\"ahler metric is
\begin{align}
	K_{s\overline{s}} &= Y^{-2}\\
	K_{s\overline{J}} &= Y^{-2}x_J^{-1}\alpha_J\\
	K_{s\overline{BJ}} &= -K_{s\overline{J}} \phi_{BJ}\\
	K_{s\overline{B}} &= -Y^{-2}P_B\widetilde{\Pi}_B\phi_B\\
	K_{I\overline{s}} &= Y^{-2}x_I^{-1}\alpha_I\\
	K_{I\overline{J}} &= \delta_{I,J}x_I^{-2}\beta_I + Y^{-2} \left(x_I x_J \right)^{-1} \alpha_I \alpha_J +\left( x_I x_J \right)^{-1}\sum\nolimits_D q_I^D q_J^D \widetilde{P}_D X_D \\
	K_{I\overline{BJ}} &= -K_{I\overline{J}} \phi_{BJ} \\
	K_{I\overline{B}} &= -x_I^{-1}q_I^B\widetilde{P}_B \widetilde{\Pi}_B \phi_B - Y^{-2}x_I^{-1} P_B \alpha_I \widetilde{\Pi}_B \phi_B\\
	K_{AI\overline{s}} &= -Y^{-2}x_I^{-1}\alpha_I\overline{\phi}_{AI} \\
	K_{AI\overline{J}} &= -K_{I\overline{J}}\overline{\phi}_{AI} \\
	K_{AI\overline{BJ}} &= K_{I\overline{J}}\overline{\phi}_{AI}\phi_{BJ} + \delta_{I,J}\delta_{A,B}x_I^{-1}\beta_I \\
	K_{AI\overline{B}} &= -K_{I\overline{B}}\overline{\phi}_{AI}\\
	K_{A\overline{s}} &= -Y^{-2}P_A\widetilde{\Pi}_A\overline{\phi}_A\\
	K_{A\overline{J}} &=  -x_Jq_J^A \widetilde{P}_A\widetilde{\Pi}_A\overline{\phi}_A - Y^{-2}x_J^{-1}P_A\alpha_J \widetilde{\Pi}_A\overline{\phi}_A\\
	K_{A\overline{BJ}} &= -K_{A\overline{J}}\phi_{BJ}\\
	K_{A\overline{B}} &= \delta_{A,B}  \widetilde{P}_A \widetilde{\Pi}_A+ Y^{-2}P_AP_B\widetilde{\Pi}_A\widetilde{\Pi}_B \overline{\phi}_A\phi_B
\end{align}
The inverse K\"ahler metric is
\begin{align}
	K^{s\overline{s}} &= Y^2+\sum\nolimits_D P_D^2 \widetilde{P}_D^{-1}X_D + \sum\nolimits_K\beta_K^{-1} b_K^2\\
	K^{s\overline{J}} &= -\beta_J^{-1}x_J b_J\\
	K^{s\overline{BJ}} &= 0\\
	K^{s\overline{B}} &= \left(P_B\widetilde{P}_B^{-1} - \sum\nolimits_K\beta_K^{-1}q_K^B b_K\right)\overline{\phi}_B\\
	K^{I\overline{s}} &= -\beta_I^{-1} x_I b_I\\
	K^{I\overline{J}} &= \delta_{I,J}\beta_I^{-1}\left(x_I \sum\nolimits_D\left|\phi_{DI}\right|^2 + x_I^2\right) \\
	K^{I\overline{BJ}} &= \delta_{I,J}\beta_I^{-1} x_I \overline{\phi}_{BI}\\
	K^{I\overline{B}} &= \beta_I^{-1}q_I^Bx_I\overline{\phi}_B \\
	K^{AI\overline{s}} &= 0 \\
	K^{AI\overline{J}} &= \delta_{I,J}\beta_I^{-1}x_I\phi_{AI}\\
	K^{AI\overline{BJ}} &= \delta_{I,J}\delta_{A,B}\beta_I^{-1}x_I\\
	K^{AI\overline{B}} &= 0\\
	K^{A\overline{s}} &= \left(P_A\widetilde{P}_A^{-1}-\sum\nolimits_K\beta_K^{-1}q_K^Ab_K\right)\phi_A\\
	K^{A\overline{J}} &= \beta_J^{-1}q_J^A x_J \phi_A\\
	K^{A\overline{BJ}} &= 0\\
	K^{A\overline{B}} &= \delta_{A,B}\widetilde{P}_A^{-1}\widetilde{\Pi}_A^{-1}+\sum\nolimits_K\beta_K^{-1}q_K^Aq_K^B\phi_A\overline{\phi}_B\\
\end{align}


\subsection{The Linear Superfield Formalism} \label{kml}

The K\"ahler potential in section \ref{lsf} is
\begin{equation}
	K = \ln(\ell) + g(\ell) - \sum\nolimits_I \ln x_I + \sum\nolimits_A X_A.
\end{equation}
We define the following quantities,
\begin{align}
	\widetilde{b} &\equiv 1+\ell b \\
	\widetilde{P}_A &\equiv 1+\ell P_A \\
	\beta_I &\equiv \widetilde{b} + \sum\nolimits_D q_I^D \widetilde{P}_D X_D  \\
	\widetilde{\Pi}_A &\equiv \Pi_I x_I^{-q_I^A}.
\end{align}
The effective K\"ahler metric is
\begin{align}
	\widehat{K}_{s\overline{s}} &= \ell^2 (1+\ell g')^{-1}\\
	\widehat{K}_{s\overline{J}} &= 0\\
	\widehat{K}_{s\overline{BJ}} &= 0\\
	\widehat{K}_{s\overline{B}} &= 0\\
	\widehat{K}_{I\overline{s}} &= 0\\
	\widehat{K}_{I\overline{J}} &= \delta_{I,J}x_I^{-2}\beta_I + (x_I x_J)^{-1}\sum\nolimits_D \widetilde{P}_D q_I^D q_J^D X_D \\
	\widehat{K}_{I\overline{BJ}} &= -K_{I\overline{J}} \phi_{BJ} \\
	\widehat{K}_{I\overline{B}} &= -x_I^{-1}q_I^B\widetilde{P}_B\widetilde{\Pi}_B \phi_B\\
	\widehat{K}_{AI\overline{s}} &= 0\\
	\widehat{K}_{AI\overline{J}} &= -\widehat{K}_{I\overline{J}}\overline{\phi}_{AI}\\
	\widehat{K}_{AI\overline{BJ}} &= \widehat{K}_{I\overline{J}} \overline{\phi}_{AI}\phi_{BJ} + \delta_{I,J}\delta_{A,B}x_I^{-1}\beta_I \\
	\widehat{K}_{AI\overline{B}} &=  -\widehat{K}_{I\overline{B}}\overline{\phi}_{AI}\\
	\widehat{K}_{A\overline{s}} &= 0\\
	\widehat{K}_{A\overline{J}} &=  -x_J^{-1}q_J^A\widetilde{P}_A\widetilde{\Pi}_A\overline{\phi}_A\\
	\widehat{K}_{A\overline{BJ}} &= -\widehat{K}_{A\overline{J}}\phi_{BJ}\\
	\widehat{K}_{A\overline{B}} &= \delta_{A,B} \widetilde{P}_A\widetilde{\Pi}_A
\end{align}
The inverse effective K\"ahler metric is
\begin{align}
	\widehat{K}^{s\overline{s}} &= \ell^{-2}(1+\ell g')\\
	\widehat{K}^{s\overline{J}} &= 0\\
	\widehat{K}^{s\overline{BJ}} &= 0\\
	\widehat{K}^{s\overline{B}} &= 0\\
	\widehat{K}^{I\overline{s}} &= 0\\
	\widehat{K}^{I\overline{J}} &= \delta_{I,J}\beta_I^{-1}\left(x_I \sum\nolimits_D\left|\phi_{DI}\right|^2 + x_I^2\right) \\
	\widehat{K}^{I\overline{BJ}} &= \delta_{I,J}\beta_I^{-1} x_I \overline{\phi}_{BJ}\\
	\widehat{K}^{I\overline{B}} &= \beta_I^{-1}q_I^Bx_I\overline{\phi}_B \\
	\widehat{K}^{AI\overline{s}} &= 0 \\
	\widehat{K}^{AI\overline{J}} &= \delta_{I,J}\beta_I^{-1}x_I\phi_{AI}\\
	\widehat{K}^{AI\overline{BJ}} &= \delta_{I,J}\delta_{A,B}\beta_I^{-1}x_I\\
	\widehat{K}^{AI\overline{B}} &= 0\\
	\widehat{K}^{A\overline{s}} &= 0\\
	\widehat{K}^{A\overline{J}} &= \beta_J^{-1}q_J^A x_J \phi_A\\
	\widehat{K}^{A\overline{BJ}} &= 0\\
	\widehat{K}^{A\overline{B}} &= \delta_{A,B}\widetilde{P}_A^{-1}\widetilde{\Pi}_A^{-1}+\sum\nolimits_K\beta_K^{-1}q_K^Aq_K^B\phi_A\overline{\phi}_B\\
\end{align}


\section{Generating VEVs in the Chiral Superfield Formalism} \label{chivevs}
In this appendix we reproduce section \ref{vevs} in the chiral superfield formalism.  The $D$-term contribution to the scalar potential is
\begin{equation} \label{chidpotgen}
	V_D=\frac{1}{2}g^2\left(\sum\nolimits_\alpha q_\alpha K_\alpha \phi_\alpha + \xi_D\right)^2,
\end{equation}
but now with the Fayet-Illiopoulos term
\begin{equation}
	\xi_D=\frac{2Y^{-1} \tn{Tr}(Q)}{192\pi^2}.
\end{equation}
With the K\"ahler potential (\ref{chikpot}), the $D$-term (\ref{chidpotgen}) becomes
\begin{equation} \label{chidpot} 
	V_D=\frac{1}{2}g^2\left[\sum\nolimits_\alpha\left(\prod\nolimits_I x_I^{-q_I^\alpha}\right)q_\alpha \xi_\alpha|\phi_\alpha|^2 + \xi_D\right]^2,
\end{equation}
where we have defined
\begin{equation}
	\xi_\alpha\equiv\left\{
		\begin{aligned}
			&Y^{-1}\left[Y+b_I+\sum\nolimits_B q_I^B X_B (Y+p_B)\right]&&\tn{for } \alpha=AI\\
			&Y^{-1}(Y+p_A) &&\tn{for } \alpha=A.
		\end{aligned} \right.
\end{equation}
Canceling (\ref{chidpot}) requires the matter fields to pick up the modular invariant VEVs,
\begin{equation}\label{chidvev}
	|\langle\phi_\alpha\rangle|^2=f_\alpha Y^{-1}\xi_\alpha^{-1} \prod\nolimits_I x_I^{q_I^\alpha}.
\end{equation}


\section{Background Calculations}

In some of the subsections in section \ref{modelbuilding} we assumed that certain fields obtained certain VEVs during inflation.  In this appendix we show explicitly under what circumstances this will happen.  To do so it proves helpful to write the scalar potential (\ref{pot}) as
\begin{equation} \label{vtdef}
	V=e^K \widetilde{V}.
\end{equation}
Then, designating derivatives with respect to the untwisted matter field, $\psi$, by a subscript, so that
\begin{equation}
	V_{\psi} = \frac{\partial V}{\partial \psi_I}, \qquad V_{\psi\psi'} = \frac{\partial^2 V}{\partial \psi_I \partial \psi'_{I'}},
\end{equation}
we have
\begin{align}
	&V_{\psi} = e^K \left( \widetilde{V}_{\psi} + \frac{2|\psi_I|}{x_I}\widetilde{V} \right) \label{vp}\\
	&V_{\psi \psi} \Bigr|_{\psi=0} = e^K \left(\widetilde{V}_{\psi \psi} + \frac{2}{x_I}\widetilde{V}\right) \biggr|_{\psi=0} \label{vpp}\\
	&V_{\psi \psi'} \Bigr|_{\psi,\psi'=0} =  e^K \widetilde{V}_{\psi \psi'} \Bigr|_{\psi,\psi'=0}. \label{vpp'}
\end{align}
For simplicity we have suppressed the $A$ index for the untwisted matter fields $\psi$, and in (\ref{vpp'}) $\psi$ and $\psi'$ are assumed to be distinct.  Finally, we will also need
\begin{equation} \label{ipsid}
	V_{I \psi} \Bigr|_{\psi=0} = e^K \left(\widetilde{V}_{I \psi} + \frac{1}{x_I}\widetilde{V}_{\psi}\right) \biggr|_{\psi=0}.
\end{equation}


\subsection{Section \ref{flatI}} \label {apenetaprob}

In this subsection we show under what circumstances $|\psi_1|=0$ is a stable minimum for the scalar potential considered in section \ref{flatI}.  We will find, in fact, that for the simplest case, it is actually a maximum.  Begin by writing the superpotential (\ref{flatsp}) as
\begin{equation}
	W=\psi_1 \Lambda, \qquad \Lambda=\eta_2^{-2}\eta_3^{-2},
\end{equation}
so that from (\ref{vtdef}) we have
\begin{equation}
\begin{aligned} 
	\widetilde{V}=(\ell g' -2) |\psi_1|^2|\Lambda|^2  &+ (1+b\ell)^{-1}|\psi_1|^2\sum_I\biggl[ |x_I\Lambda_I-\Lambda|^2 + x_I \sum_{A \neq \psi} |\bar{\phi}_{AI}\Lambda_I|^2\biggr] \\
	& + (1+b\ell)^{-1}x_{1} \left||\psi_1|^2 \Lambda_1+\Lambda \right|^2.
\end{aligned}
\end{equation}
We can see that $\psi_1$ only enters $\widetilde{V}$ in the form of $|\psi_1|^2$ and therefore $|\psi_1|=0$ is a critical point of $\widetilde{V}$.  From a look at (\ref{vp}) we can immediately see that $|\psi_1|=0$ is then a critical point of the full potential $V$.  This also means that (\ref{ipsid}) vanishes and so the determination of stable minima for $\psi_1$ and the K\"ahler moduli, $t_I$, may be done independently.  It follows then that $t_2$, $t_3=\exp(i\pi/6)$ correspond to minima, as discussed in section \ref{flatI}.

$|\psi_1|=0$ will be a stable minimum if (\ref{vpp}) is positive.  We have
\begin{equation} \label{Vat0}
	\left.\frac{2\widetilde{V}}{x_1}\right| _{\psi=0} = \frac{2}{1+b\ell}|\Lambda|^2,
\end{equation}
and
\begin{equation}
\begin{aligned} \label{V''at0}
	\widetilde{V}_{\psi\psi} \Bigr|_{\psi=0} = (\ell g'-2)|\Lambda|^{2} &+ (1+b\ell)^{-1}\sum_I \left[ \left| x_I \Lambda_I - \Lambda\right|^2 + x_I \sum_{A\neq \psi} \left|\bar{\phi}_{AI} \Lambda_I\right|^2\right] \\
	&+(1+b\ell)^{-1}\left[2x_1\tn{Re}\left(\Lambda_1 \overline{\Lambda} \right) - (1+b\ell)^{-1}|\Lambda|^{2}\right].
\end{aligned}
\end{equation}
Now,
\begin{equation} \label{selfdual}
	\left|x_I \Lambda_I - \Lambda\right|^2 = |\Lambda|^2 \left| 2 x_I\xi_I + 1 \right|^2, \qquad \xi_I=\frac{1}{\eta_I}\frac{\partial \eta_I}{\partial t_I},
\end{equation}
for $I=2$, 3 and $\left| 2 x_I\xi_I + 1 \right|$ vanishes at the self dual point $t_I=\exp(i\pi/6)$.  In the second term of (\ref{V''at0}) then, only the $I=1$ contribution survives.  The second to last term vanishes since $\Lambda_1=0$ and we make the usual assumption that the VEVs of the matter fields are zero (i.e. we ignore the matter fields we are not using) so that the $\phi_{AI}$ are zero.

Putting everything together by placing (\ref{Vat0}) and (\ref{V''at0}) into (\ref{vpp}) we obtain
\begin{equation}
	V_{\psi\psi}\Bigr| _{\psi=0}= e^K |\Lambda|^2\left[\ell g'-2\left(1-\frac{1}{1+b\ell}\right)\right].
\end{equation}
Using the results of section (\ref{dilstab}) we can determine numerically that when the dilaton is sitting at its minimum (during inflation),
\begin{equation}
	\left.\ell g'\right|_{\langle\ell\,\rangle}=-0.75, \qquad \langle\ell \rangle = 0.87.
\end{equation}
Thus,
\begin{equation}
			V_{\psi\psi}\Bigr| _{\psi=0}= -(1.25) e^K |\Lambda|^2,
\end{equation}
which corresponds to $|\psi_1|=0$ being an unstable maximum.  We have used $b=30/8\pi^2$ but note that even for $b=0$ (i.e. removing the Green-Schwarz counterterm) we have a maximum.

By turning on the $\phi_{AI}$ it may be possible to turn this into a minimum.  We made the usual assumption that the $\phi_{AI}$ (that were not the inflaton) were zero during inflation.  If instead we turn them on (say, by giving them $D$-term VEVs) then they contribute positively to $V_{\psi\psi}$.  Each $\phi_{AI}$ contributes
\begin{equation}
	\Delta V_{\psi\psi} \Bigr|_{\psi=0} = e^K	|F|^2 \biggl[ \frac{4}{1+b\ell}x_I|\xi_I|^2\biggr] |\phi_{AI}|^2 =  (0.431) |\phi_{AI}|^2 e^K	|\Lambda|^2.
\end{equation}
Thus, we require
\begin{equation} \label{bound}
	\sum\nolimits_{AI} (0.43)|\phi_{AI}|^2 > 1.25 \quad\Rightarrow\quad \sum\nolimits_{AI}|\phi_{AI}|^2 > 2.91.
\end{equation}

Recall from (\ref{dvev}) that for $D$-term VEVs,
\begin{equation}
	\sum\nolimits_{IA} \langle|\phi_{IA}|^2\rangle \approx \frac{2\langle \ell \rangle \times O(100) } {192\pi^2 q}x_I \geq \frac{0.16}{q}.
\end{equation}
For simplicity we have assumed each field has the an identical charge, $q$, and the inequality follows because we have evaluated the $O(100)$ term at precisely 100.  For $q$ a little smaller than 1 and $O(100)$ a little greater than 100 this can presumably satisfy the bound in (\ref{bound}).

Finally, note that giving these terms $D$-term VEVs has no effect on the potential since once we set $|\psi_1|=0$ their contribution vanishes.


\subsection{Section \ref{dinvevs}}
\label{appdinvevs}

In this subsection we argue that $|\psi_1|=|\psi'_1|=0$ corresponds to a stable minimum for the superpotential given in (\ref{wwpsp}).  We begin by writing the superpotential as
\begin{equation}
	w=\psi_1 \Lambda, \qquad w'=\psi_1' \Lambda',
\end{equation}
where
\begin{equation} \label{F's}
	\Lambda=\chi_2 \phi_2 \phi_3 \eta_2^2, \qquad \Lambda'=\phi_2' \phi_3' \sum\nolimits_n c_n \Gamma^n, \qquad \Gamma=\chi_2 \phi_2' \phi_3' \eta_2^4 \eta_3^2.
\end{equation}
Then
\begin{equation} \label{vtfvevfg}
\begin{aligned} 
	\widetilde{V}=&(\ell g' -2) |\psi_1 \Lambda + \psi_1' \Lambda'|^2 \\
	  &+ (1+b\ell)^{-1}\sum_I\biggl[ \Bigl|\psi_1(x_I \Lambda_I-\Lambda) + \psi_1'(x_I \Lambda_I'-\Lambda')\Bigr|^2 \\
	  &\qquad+ x_I \sum_{A \neq \psi,\psi'} \Bigl|\psi_1(\bar{\phi}_{AI}\Lambda_I + \Lambda_{AI})  + \psi_1'(\bar{\phi}_{AI}\Lambda_I' + \Lambda_{AI}') \Bigr|^2\biggr] \\
	& + (1+b\ell)^{-1}x_1 \biggl[ \Bigl|\psi_1(\bar{\psi}_1 \Lambda_I) + \psi_1'(\bar{\psi}_1 \Lambda_I') + \Lambda\bigr|^2 + \Bigl| \psi_1(\bar{\psi}_1' \Lambda_I) + \psi_1'(\bar{\psi}_1' \Lambda_I') + \Lambda'\Bigr|^2 \biggr].
\end{aligned}
\end{equation}
It is easy to see that $|\psi_1|$ and $|\psi_1'|$ never appear alone and raised to the first power, so that $|\psi_1|=|\psi_1|=0$ are critical points of $\widetilde{V}$ and thus from (\ref{vp}) also critical points of $V$.  This also means that (\ref{ipsid}) vanishes and so the determination of stable minima for $\psi_1$ and the K\"ahler moduli, $t_I$, may be done independently.

Note that if the $\psi$ fields are at these points then the potential becomes
\begin{equation}
	V=e^K(1+b\ell)^{-1}x_1 (|\Lambda|^2+|\Lambda'|^2).
\end{equation}
As discussed in section \ref{dinvevs} we take the $c_n$ to be such that the potential is minimized for
\begin{equation}
	\langle \Lambda' \rangle = 0.
\end{equation}

To determine if our critical points correspond to minima we use the second derivative test.  For two fields this requires that both
\begin{equation} \label{Deq}
	D \equiv V_{\psi \psi} V_{\psi' \psi'} - V_{\psi \psi'}^2
\end{equation}
and $V_{\psi \psi}$ are positive (which itself requires $V_{\psi' \psi'}$ to be positive).  Evaluated at the critical points, $|\psi_1|=|\psi_1'|=0$, we have
\begin{equation}
	\left. V_{\psi\psi}\right| = \left. e^K\left(\widetilde {V}_{\psi\psi} + \frac{2\widetilde{V}}{x_1}\right)\right|, \qquad \left. V_{\psi'\psi'}\right| = \left. e^K\left(\widetilde {V}_{\psi'\psi'} + \frac{2\widetilde{V}}{x_1}\right)\right|, \qquad \left. V_{\psi\psi'}\right| = \left. e^K\widetilde {V}_{\psi\psi'} \right|,
\end{equation}
where the bar means that we have set the $\psi$ fields to zero and where
\begin{equation}
	\left.\frac{2\widetilde{V}}{x_1}\right| = \frac{2}{1+b\ell}\left(|\Lambda|^2+|\Lambda'|^2\right)=\frac{2}{1+b\ell}|\Lambda|^2.
\end{equation}

Of the second derivatives the simplest is
\begin{equation} \label{psippsip}
	\left. e^{-K} V_{\psi' \psi'}\right| = (1+b\ell)^{-1} \left[|x_I \Lambda_I'|^2 + x_I\sum_{A\neq \psi, \psi'} | \bar{\phi}_{AI} \Lambda'_I + \Lambda'_{AI} |^2 + |\Lambda|^2 \right],
\end{equation}
where we have used $\langle \Lambda' \rangle =0$.  We can clearly see that this is positive.  

Now consider
\begin{equation}
\begin{aligned}
	\left. e^{-K} V_{\psi \psi}\right| = &(\ell g' -2)^{-1}|\Lambda|^2 + (1+b\ell)^{-1} \sum_I \left[ |x_I \Lambda_I - \Lambda|^2 + x_I \sum_{A\neq \psi, \psi'} | \bar{\phi}_{AI}\Lambda_I + \Lambda_{AI} |^2 \right] \\
	& + (1+b\ell)^{-1}\left[ 2x_1 \tn{Re} (\Lambda_1 \bar{\Lambda}) + |\Lambda|^2\right].
\end{aligned}
\end{equation}
There are a few things to note about this equation.  First, we have again used $\langle \Lambda' \rangle =0$.  Second the $\tn{Re} (\Lambda_1 \bar{\Lambda})$ vanishes since $\Lambda$ is independent of $t_1$.  This is a direct consequence of our choice for $\chi_2$ to be in the $I=2$ sector and is a major reason why such a choice was made.  In fact, $\Lambda_I$ vanishes except for when $I=2$.  This means that the second term only vanishes when $I=2$ in the sum, since $t_2=\exp(i\pi/6)$ (see (\ref{selfdual})).  Putting it all together we have
\begin{equation} \label{psipsid}
		\left. e^{-K} V_{\psi \psi}\right| = |\Lambda|^2\left[ \ell g' - \left(2-\frac{3}{1+b\ell} \right) \right] + \frac{1}{1+b\ell} \sum_I x_I \sum_{A\neq \psi, \psi'} | \bar{\phi}_{AI}\Lambda_I + \Lambda_{AI} |^2.
\end{equation}
The first term is negative (though not as negative as in section \ref{apenetaprob} because we now have a 3 instead of a 2).  In section \ref{apenetaprob} we had to turn on VEVs to get the analogous second derivative positive.  Here we don't have a choice since we already have fields with $D$-term VEVs.  Specifically, we have
\begin{equation} \label{negcoef}
	\ell g' - \left(2-\frac{3}{1+b\ell} \right) = -0.50
\end{equation}
and
\begin{equation}
\begin{aligned}
	\frac{1}{1+b\ell}&\sum_I x_I \sum_{A\neq \psi, \psi'} | \bar{\phi}_{AI}\Lambda_I + \Lambda_{AI} |^2  \\
	&= (1+b\ell)^{-1}|\Lambda|^2\left[x_2\left(|2\bar{\psi}_2\zeta_2 + \psi_2^{-1}|^2 + |2\bar{\phi}_2 \zeta_2 + \phi_2^{-1} |^2 + |2\bar{\phi_2}'\zeta_2|^2 \right) + x_3|\phi_3|^{-2}\right] \\
	& \apprge (1+b\ell)^{-1} x_3|\phi_3|^{-2}|\Lambda|^2 \sim (0.75)(1.7)(0.1)^{-2}|\Lambda|^2 \sim 100|\Lambda|^2.
\end{aligned}
\end{equation}
Thus, (\ref{psipsid}) is easily positive.  This follows since $\Lambda_{AI}$ leads to a large contribution.  Such a contribution did not show up in section \ref{apenetaprob} since there the superpotential was linear in the matter fields and so $\Lambda_{AI}$ vanished.

The final second derivative is
\begin{equation} \label{crossterm}
\begin{aligned}
	 e^{-K} & V_{\psi \psi'} \Bigr| = 2(1+b\ell)^{-1} \tn{Re}\Biggl\{ e^{i(\theta-\theta')} \\
	&\times \Biggl[ \sum_I \biggl\{ 2(x_I\Lambda_I-\Lambda) x_I\bar{\Lambda}_I' + x_I\sum_{A\neq \psi,\psi'} (\bar{\phi}_{AI} \Lambda_I + \Lambda_{AI}) (\phi_{AI}\bar{\Lambda}_I' + \bar{\Lambda}_{AI}') \biggr\} + \Lambda\bar{\Lambda}_I' \Biggr]\Biggr\}.
\end{aligned}
\end{equation}
We have so far shown that $V_{\psi\psi}$ and $V_{\psi'\psi'}$ are positive.  We now argue that $V_{\psi\psi'}$ is smaller than them, and therefore that (\ref{Deq}) is positive.  From (\ref{F's}) it is not hard to show that $\Lambda'_{AI}$ to leading order is order four in the matter fields.  The dominant contribution (i.e. the lowest leading order term in the matter fields) to (\ref{crossterm}) is the $\Lambda_{AI}\overline{\Lambda}'_{AI}$ which is order six.  However, the dominant term to (\ref{psipsid}) is the $|\Lambda_{AI}|^2$ term which is order four and the dominant term to (\ref{psippsip}) is the $|\Lambda|^2$ term which is order six.  We therefore expect (\ref{Deq}) to be positive and $|\psi_1|=|\psi'_1|=0$ to correspond to a stable minimum.


\subsection{Section \ref{loopmod1}}
\label{apploopmod1}

Here we are considering a potential of the form
\begin{equation} \label{apensploop}
	W=\phi_1\phi''_2\phi''_3 + W_0,
\end{equation}
where $W_0$ is the superpotential in section \ref{dinvevs}.  We would like to show that the VEVs of the $\phi$ fields in the first term of (\ref{apensploop}) and the $\psi$ fields in $W_0$ vanish.  First note that, under the assumption that a critical point of the potential exists when the fields vanish, it is not hard to show that the analysis for the minima of the $\phi$ fields is independent of the analysis for the $\psi$ fields.

If we set the $\phi$ fields in the potential to zero then it reduces to the same potential that is in section \ref{dinvevs}, where VEVs of the $\psi$ fields were argued to vanish.  Now set the $\psi$ fields at their (vanishing) VEVs, but retain the $\phi$ fields, to get
\begin{equation}
\begin{aligned}
	e^{-K}V=&|\phi_1\phi_2\phi_3|^2\left[ \ell g' - 2 \left(1 - \frac{3}{1+b\ell} \right)\right] \\
	&+ \frac{1}{1+b\ell} \left( x_1|\phi''_2\phi''_3|^2+ x_2|\phi_1\phi''_3|^2 + x_3|\phi_1\phi''_2|^2 + |\Lambda|^2\right),
\end{aligned}
\end{equation}
where $\Lambda$ is given in (\ref{F's}).  With (\ref{negcoef}) we know that the first term is negative.  If it were positive then the entire potential would be positive and the VEVs would be zero.  However, it is the terms in the second line that dominate, being fourth order in the matter fields while the first term is order six.  These terms tell us that for vanishing values of the fields we have a minimum.


\end{document}